\documentclass[twoside]{ilcws07}
\usepackage[latin1]{inputenc}
\usepackage[dvips]{graphicx,epsfig,color}
\usepackage{wrapfig,rotating}
\usepackage{amssymb,amsmath,array}

\begin{document}
\title{
New physics effect on the top-Yukawa coupling at ILC} 
\author{Shinya~Kanemura$^1$, Koichi~Matsuda$^2$, Daisuke~Nomura$^3$, and Koji~Tsumura$^3$\footnote{
Presented at the LCWS07, DESY, Hamburg
}
\thanks{K. T. was supported, in part, by the Grant-in-Aid of the Ministry of Education, Culture, Sports, Science and Technology, Government of Japan, Grant No. 16081207.}
\vspace{.3cm}\\
1- Department of Physics, University of Toyama\\
3190 Gofuku, Toyama 930-8555, Japan\\
\vspace{.1cm}\\
2- Center for High Energy Physics, Tsinghua University\\
Beijing, 100084, China\\
\vspace{.1cm}\\
3- Theory Group, KEK\\
1-1 Oho, Tsukuba 305-0801, Japan\\
}

\maketitle

\begin{abstract}
Measurement of the top-Yukawa coupling is important to understand the fermion mass generation mechanism and dynamics of electroweak symmetry breaking. We discuss the top quark anomalous couplings which can be described by higher dimensional operators. We investigate the process $e^-e^+ \to  W^-W^+\nu\bar\nu \to t \bar t \nu\bar\nu$ to study the contribution of the anomalous top-Higgs coupling to the cross section. The effect of the dimension-six top-Higgs interaction on the cross section can be a few hundred percent greater than the SM prediction. Such a large effect can be measured at the International Linear Collider. 
\end{abstract}

\section{Introduction}
The mass of the top quark has been measured to be at the scale of the electroweak symmetry breaking, so that the top-Yukawa coupling has turned out to be of order one in the standard model (SM). It is a quite natural scale as compared to the other quarks. This fact would indicate a relation between the top quark physics and the dynamics of electroweak symmetry breaking. Top quark motivated models such as the top mode condensation, top color models, and top flavor models have been discussed in literature. These models generally predict rather strong dynamics for the electroweak symmetry breaking. Measuring the top-Yukawa coupling is essentially important not only to confirm the SM but also to test new physics models including them.

\begin{figure}
\begin{center}
\includegraphics[width=5.0cm]{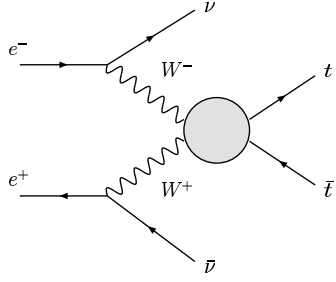}
\caption{
The top pair production via  $W$ boson fusion 
}
\label{fig:diagram-full}
\end{center}

\end{figure}

Information of Higgs coupling constants can be extracted at future experiments. However, measurement of the top-Yukawa coupling may be difficult because of the huge QCD backgrounds. Determination of the coupling constants can be performed at the International Linear Collider (ILC). At the ILC, the top-Yukawa interaction is expected to be measured through the process $e^-e^+ \to t \bar t H$\cite{tth} for a relatively light Higgs boson when it is kinematically allowed. For a heavier Higgs boson, it would be detectable via the vector boson fusion process $e^-e^+ \to  W^-W^+\nu\bar\nu \to t \bar t \nu\bar\nu$\cite{WBF-tt} in Fig.\ref{fig:diagram-full}.

\section{Dimension-six top quark operators}
At low energies, the non-SM interaction can be expressed by using the higher dimensional operators. Such operators are introduced by integrating out the heavy new physics particles. In this section, we study the dimension-six operators which affect the top quark interaction.

Below the SM cutoff scale $\Lambda$, the new physics effect which is related to the top quark can be described by the effective Lagrangian as
	\begin{align}
	{\mathcal L}_\mathrm{eff}^{}
	= {\mathcal L}_\mathrm{SM}^{}+{\mathcal L}_\mathrm{dim.6}^{}
	+{\mathcal L}_\mathrm{dim.8}^{}+\cdots,
	\end{align}
where ${\mathcal L}_\mathrm{SM}^{}$ is the SM Lagrangian, and 
	\begin{align}
	{\mathcal L}_{\mathrm{dim.}^{}n}= \frac1{\Lambda^{n-4}_{}}
	\sum_iC_i^{}{\mathcal O}_i^{(n)},\quad(n\ge6),
	\end{align}
where ${\mathcal O}_i^{(n)}$ are $SU(2)_L^{}\times U(1)_Y^{}$ gauge invariant dimension-$n$ operators, and $C_i^{}$ are the coupling strength of ${\mathcal O}_i^{(n)}$. In this talk, we treat only the dimension-six operators because they should give the new physics interaction. 

The dimension-six operators have been already discussed in literature. All the gauge invariant operators are given in Refs.~\cite{buchmuller,gounaris1}. We concentrate on the CP-conserving top quark operators whose coefficients are real, 
	\begin{align}
	&{\mathcal O}_{t1}^{}
	= \left(\Phi^\dag_{}\Phi-\frac{v^2_{}}2\right)\left(\bar{q}_L^{}
	t_R^{}\tilde{\Phi}+\mathrm{H.c.}\right),\nonumber\\
	&{\mathcal O}_{t2}^{}
	= i\left(\Phi^\dag_{}D_\mu^{}\Phi\right)\bar{t}_R^{}
	\gamma^\mu_{}t_R^{}+\mathrm{H.c.},\nonumber\\
	&{\mathcal O}_{t3}^{}
	= i\left({\tilde\Phi}^\dag_{}D_\mu^{}\Phi\right)\bar{t}_R^{}
	\gamma^\mu_{}b_R^{}+\mathrm{H.c.},\\
	&{\mathcal O}_{Dt}^{}
	= \left(\bar{q}_L^{}D_\mu^{}t_R^{}\right)\left(D^\mu_{}\tilde{\Phi}
	\right)+\mathrm{H.c.},\nonumber\\
	&{\mathcal O}_{tW\Phi}^{}
	= \left(\bar{q}_L^{}\sigma^{\mu\nu}_{}{\vec \tau}t_R^{}\right)
	\tilde{\Phi}{\vec W}_{\mu\nu}^{}+\mathrm{H.c.},\nonumber\\
	&{\mathcal O}_{tB\Phi}^{}
	= \left(\bar{q}_L^{}\sigma^{\mu\nu}_{}t_R^{}\right)
	\tilde{\Phi}B_{\mu\nu}^{}+\mathrm{H.c.},\nonumber
	\end{align}
where $q_L^{}=\left(t_L^{},b_L^{}\right)^T_{}$, $\Phi$ is the scalar isospin doublet (the Higgs doublet) with hypercharge $Y=1/2$, and $\tilde{\Phi} \equiv i\,\tau_2^{}\Phi^\ast_{}$ with $\tau_i^{}$ ($i=1$--$3$) being the Pauli matrices. The dimension-six operators for the tau-Yukawa coupling have been discussed in Ref.~\cite{trott}.

Because we have not measured any Higgs coupling yet, the size of the anomalous coupling $C_{t1}^{}$ is completely free from constraints from the experimental data. The top quark gauge interaction can be directly constrained by the Tevatron, but there are no severe bounds due to lower statistics. 
From the indirect measurement, the LEP precision results can give a stringent limit to the anomalous couplings $C_i^{}\cite{gounaris2}$. The coupling $C_{Dt}^{}$ is subject to weaker bound than the others. We here set the coupling strength of ${\mathcal O}_{t2}^{},{\mathcal O}_{t3}^{},{\mathcal O}_{tW\Phi}^{},{\mathcal O}_{tB\Phi}^{}$ to be zero. 
The couplings $C_i$ of the dimension-six operators can be constrained by perturbative unitarity\cite{LQT}. Their bounds for $C_{t1}^{}$ and $C_{Dt}^{}$ are evaluated as\cite{gounaris2}
\begin{align}
&\left|C_{t1}^{}\right|\le \frac{16\pi}{3\sqrt2}\left(\frac{\Lambda}{v}
\right),\\
&-6.2\le C_{Dt}^{}\le 10.2,
\end{align}
where $v\left(\simeq 246\mathrm{GeV}\right)$ is the vacuum expectation value of the Higgs boson. 

\section{New Physics effect on the top-Yukawa coupling}

In this section, we study the phenomenology of the new physics effect in terms of the dimension-six operators. Han et al. have discussed those on the process of $e^-e^+ \to t \bar t H$\cite{han}. Those effects on the W-boson fusion $e^-e^+ \to W^-W^+ \nu \bar \nu \to t \bar t \nu \bar \nu$ are discussed in Ref.~\cite{KNT}. 

The W-boson fusion process has been studied in the SM\cite{yuan,godfrey}, and its QCD correction has also been studied in Ref.~\cite{godfrey2}. In the SM without the Higgs boson, instead of including the dimension-five operators, this process has been investigated in Ref.~\cite{larios}.  

\begin{figure}[h]
\begin{center}
\rotatebox{0}{\includegraphics[width=5.5cm]{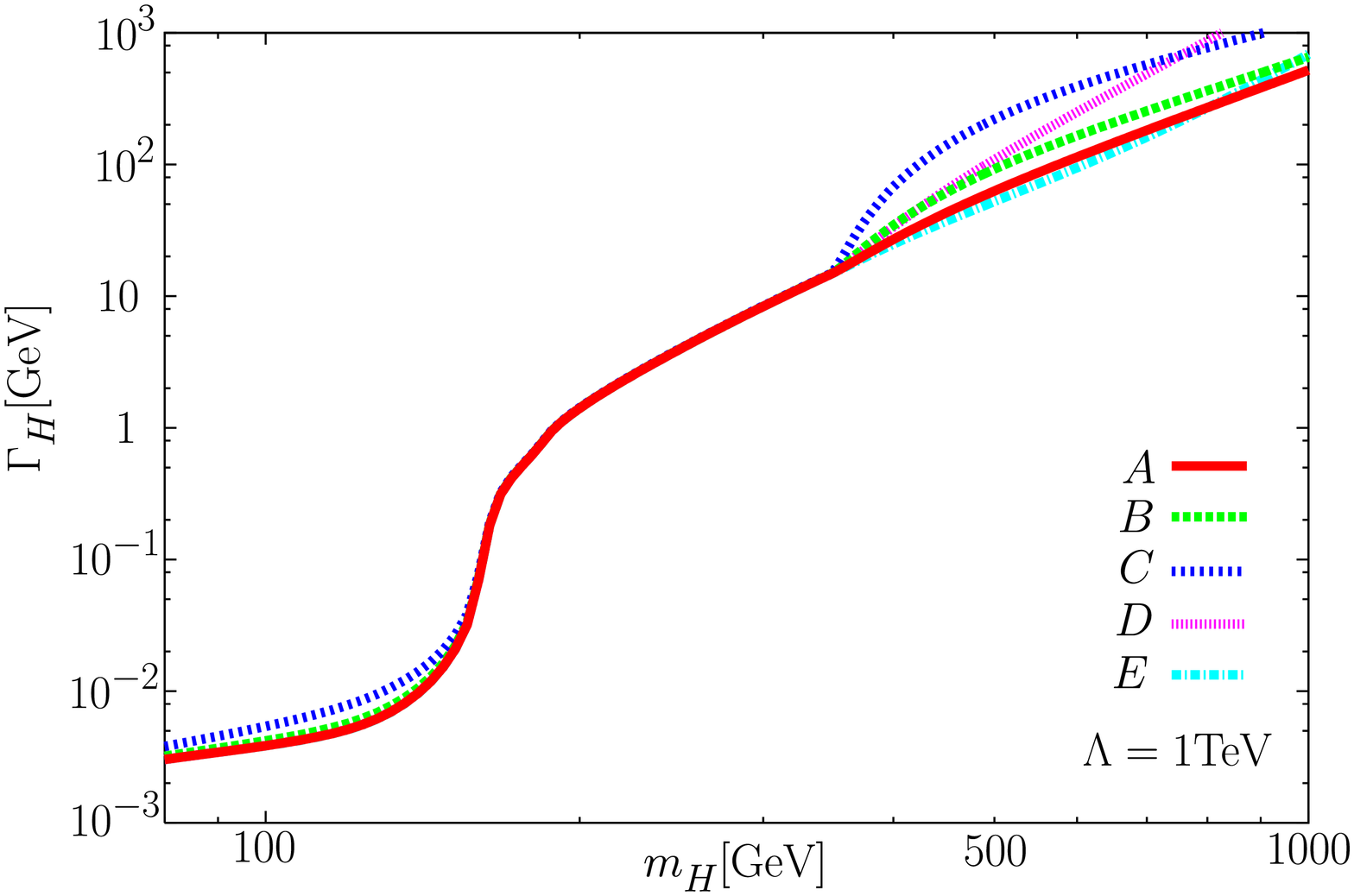}}
\caption{
The total width of the Higgs boson for several   
cases of $C_{t1}^{}$ and $C_{Dt}^{}$
. $\Lambda$ is set to be 1 TeV. 
}
\label{fig:plot-width}
\end{center}
\end{figure}

Since we have introduced the dimension-six top-Higgs interaction, the partial decay width for the process $H\to t\bar t$ is modified at tree level, which is obtained by using the effective top-Yukawa coupling 
\begin{align}
y^{\rm eff}_t(q^2,\Lambda) =
 y_t^{\rm SM}
 - v^2 \frac{C_{t1}}{\Lambda^2} 
- q^2 \frac{C_{Dt}}{2 \Lambda^2}. 
\label{eq:yeff}
\end{align}
The loop induced decay widths of $H\to \gamma\gamma, Z\gamma, gg$ are also given by replacing $y_t^{\rm SM}$ by $y_t^{\rm eff}(m_H^2)$ in the corresponding SM expressions. 
In Fig.~\ref{fig:plot-width}, the values of the total width of the Higgs boson are shown for each set of the dimension-six couplings in Tab.~\ref{tab-set} , according to the unitarity bounds. 
\begin{table}
\begin{center}
\begin{tabular}{|c||c|c|c|c|c|}
\hline
$C_i$&Set A & Set B & Set C & Set D & Set E\\
\hline \hline
$C_{t1}$&$0$&$-\frac{16\pi}{3\sqrt2}\frac{\Lambda}v$&$+\frac{16\pi}{3\sqrt2}\frac{\Lambda}v$&$0$&$0$\\
$C_{Dt}$&$0$&$0$&$0$&$+10.2$&$-6.2$\\
\hline
\end{tabular}
\caption{Sets of the dimension-six couplings we used for 
the analyses.}
\label{tab-set}
\end{center}
\end{table}
Set A corresponds to the SM case. The decay modes $H\to t\bar t$ (tree), $H \to \gamma\gamma$, $H \to \gamma Z$ and  $H \to gg$ (one loop) are largely modified at the leading order by the inclusion of ${\cal O}_{t1}$ and ${\cal O}_{Dt}$.

We here evaluate the cross section of the full process 
$e^-e^+ \to W^-W^+ \nu \bar \nu \to t \bar t \nu \bar \nu$ 
in the method of the effective W-boson approximation (EWA)\cite{ewa} 
by combining the result of calculation of the subprocess.  
We also evaluated the cross section by the full matrix element calculation by using CalcHEP\cite{calchep} (and LanHEP\cite{lanhep}), and compared the consistency with the EWA results. The EWA gives reasonable results for a large value of $\sqrt{\hat{s}}$ 
as compared to $m_W^{}$. In order to keep the validity of the calculation 
based on the EWA, we need to make the kinematic cut at an appropriate 
value. Here we employ the cut $M_{tt}^{}>400$ GeV\cite{godfrey}.
The accuracy of the EWA has been discussed in Ref.~\cite{Johnson:1987tj}. Our results agree with those in  Ref.~\cite{larios} where
expected error is evaluated to be of the order of 10\% for the cut 
$M_{tt}^{}>500$ GeV.

\begin{figure}[h]
\begin{minipage}{7.5cm}
\unitlength=1cm
\rotatebox{0}{
\includegraphics[width=6.5cm]{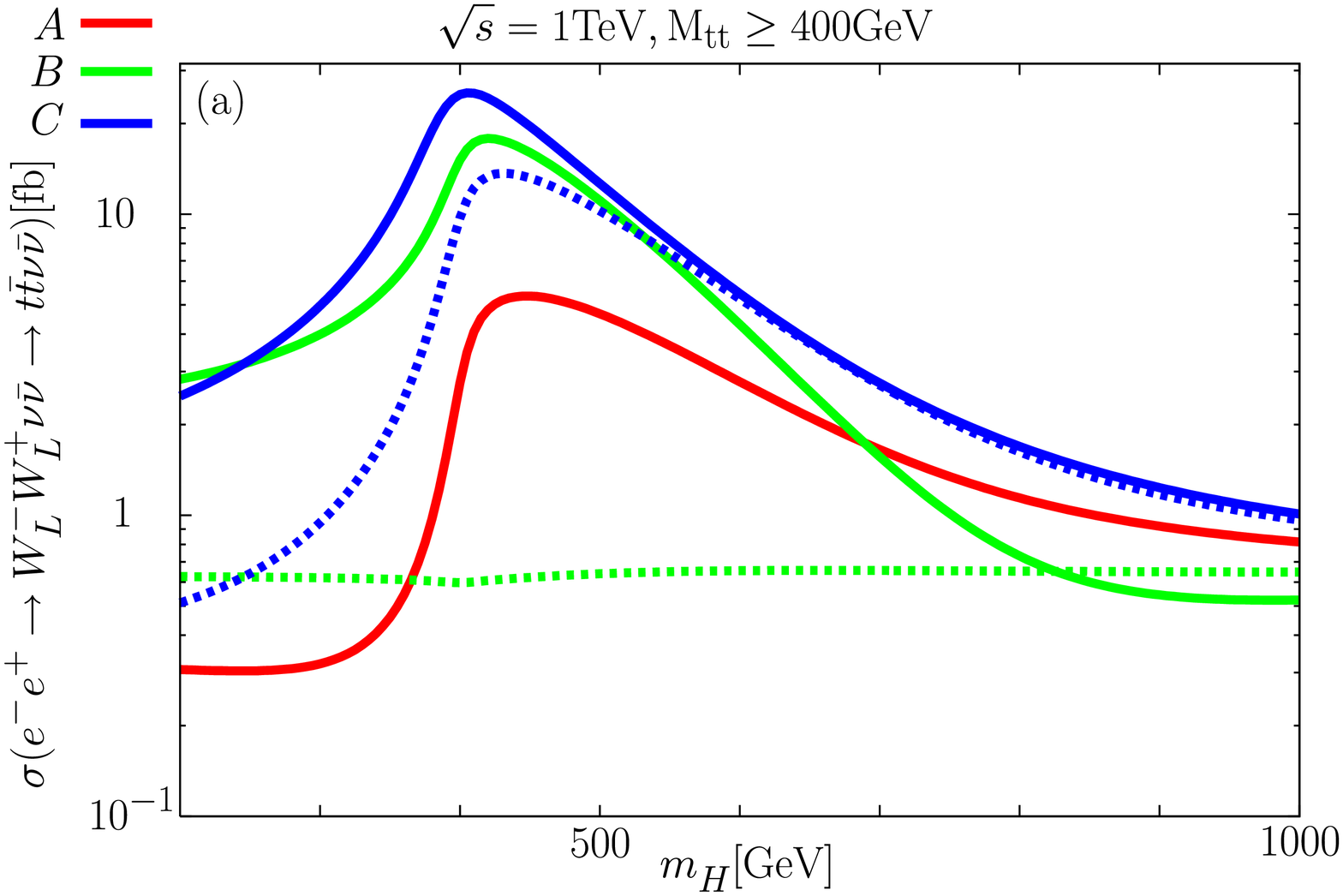}}
\end{minipage}
\begin{minipage}{7.5cm}
\unitlength=1cm
\rotatebox{0}{
\includegraphics[width=6.5cm]{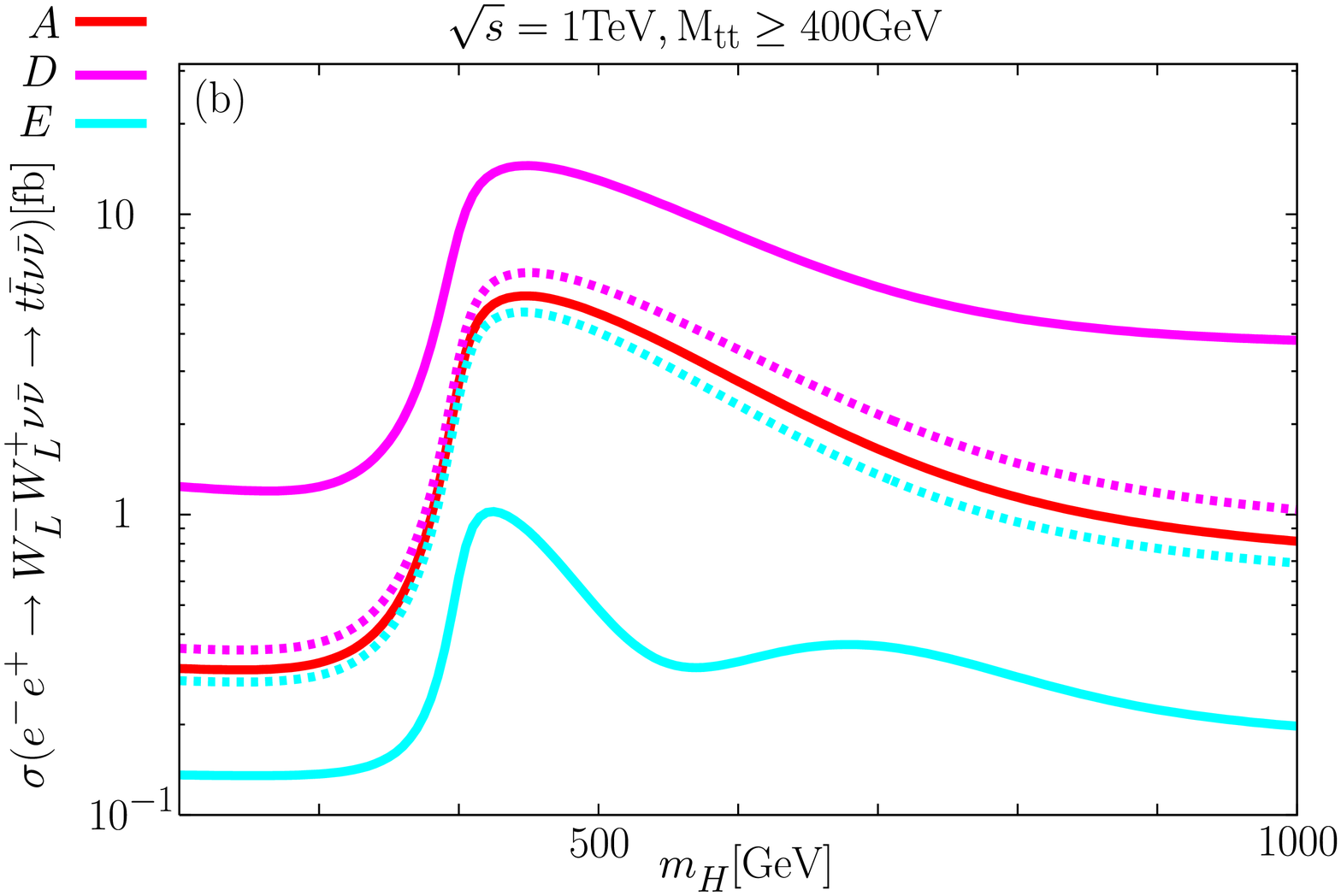}}
\end{minipage}
\caption{Cross sections of
 $e^-e^+\to W^-_L W^+_L \nu \bar \nu \to t \bar t \nu \bar \nu$
 are shown as a function of Higgs boson mass for the cases of
 Set A, Set B and Set C [Fig.~\ref{fig:SigmaCUT400-t1}(a)], and
 for those of Set A, Set D and Set E [Fig.~\ref{fig:SigmaCUT400-t1}(b)]
 with $\Lambda=1$ TeV (solid curves) and $3$ TeV (dashed curves).
 The collider energy is taken to be $\sqrt{s}=1$ TeV.
}
\label{fig:SigmaCUT400-t1}
\end{figure}

We add the dimension-six operators 
${\cal O}_{t1}$ and ${\cal O}_{Dt}$ to the SM Lagrangian. 
In Figs.~\ref{fig:SigmaCUT400-t1}(a)  and \ref{fig:SigmaCUT400-t1}(b)  
  cross sections for $e^- e^+ \to W^-_L W^+_L
 \nu \bar \nu \to  t \bar{t} \nu \bar{\nu}$ 
 are shown as a function of $m_H^{}$
 after the kinematic cut $M_{tt} \ge 400$ GeV.
 The collision energy is set to be $\sqrt{s}=1$ TeV.
 The new physics scale $\Lambda$ is assumed to be $1$ TeV and $3$ TeV.
 Fig.~\ref{fig:SigmaCUT400-t1}(a) shows 
 the results for Set B and Set C, and  
 Fig.~\ref{fig:SigmaCUT400-t1}(b) does those for Set D and Set E. 
 In both figures, the result in the SM case [Set A] is also
 plotted. 
%
%

 The SM value of the cross section for $e^-e^+ \to W^-W^+ \nu \bar \nu
 \to t \bar t \nu \bar \nu$ is order $1$fb for heavy Higgs bosons
 ($m_{H}^{} \gtrsim 400$ GeV).  
 At the ILC with an $e^-e^+$ energy of 1 TeV and
 the integrated luminosity of 500fb$^{-1}$, over several hundred events 
 are produced. 
 Naively, the statistic error of the cross section measurement
 can be less than  about 10\% level.
 The QCD corrections are evaluated to be the same order of magnitude\cite{godfrey}.
 Therefore, we can expect that the large correction of the cross section can easily be observed as long as it changes the cross section by a few times 10\% or more. 
 The effect of ${\cal O}_{Dt}$ under the constraint from the LEP data
 may also be observed when it changes the SM cross section by
 10-20 \%.
 
Background processes are also taken into account.
 Main background is 
 $e^-e^+\to \gamma t\bar t$ with $\gamma$ to be missed.
 It can be reducible by making a kinematic cut for the transverse 
 momentum of the final top quark. 
 In Ref.~\cite{larios}, the simulation study for the background reduction
 has been performed in the SM, and 
 the background can be sufficiently suppressed by the kinematic cuts.
 Another important background is the top pair production process
 via the photon fusion $\gamma\gamma\to t \bar t$.
 This mode can be suppressed by the cut $E\!\!\!\!/ > 50$
 GeV\cite{WBF-tt}, where $E\!\!\!\!/$ is the missing energy.
 Finally, the direct top-pair production $e^-e^+ \to t \bar t$
 can be suppressed by imposing the cut for the invariant
 mass $M_{tt}$.

\section{Conclusions}
In this talk, we have studied the new physics effect of the dimension-six top quark operators. Theoretical and experimental constraints on these operators have been discussed. We have evaluated the cross section of the process $e^-e^+ \to W^-W^+ \nu \bar \nu \to t \bar t \nu \bar \nu$ in the SM with the dimension-six top-Higgs interaction, and found that the deviation from the SM result can be a few hundred percent greater than the SM one, which can easily be detected at a future linear collider including the ILC. Such a large deviation may also be detectable at the LHC via the process such as $pp \to W^-W^+ X\to t \bar t X$ even though the QCD background is huge. Detailed simulation study should be necessary to clarify the significance. 

\begin{footnotesize}



%

\end{footnotesize}


\end{document}